\documentclass[preprint2]{aastex}
\begin{document}

\title{The velocity field of baryonic gas in the universe}

\author{Bryan Kim\altaffilmark{1}, Ping He\altaffilmark{2,3},
  Jes\'us Pando\altaffilmark{1}, Long-Long Feng\altaffilmark{2,4} and
  Li-Zhi Fang\altaffilmark{3}}

\altaffiltext{1}{Department of Physics, DePaul University, Chicago
Il, 60614}
\altaffiltext{2}{National
Astronomical Observatories, Chinese Academy of Science, Chao-Yang
District, Beijing, 100012, P.R. China}
\altaffiltext{3}{Department
of Physics, University of Arizona, Tucson, AZ 85721}
\altaffiltext{4}{Purple Mountain Observatory, Chinese
Academy of Sciences, Nanjing, 210008}

\begin{abstract}

The dynamic evolution of the baryonic intergalactic
medium (IGM) caused by the underlying dark matter gravity is
governed by
the Navier-Stokes equations in which many cooling and heating
processes are involved. However, it has long been
recognized that the growth mode dynamics of cosmic matter
clustering
can be sketched by a random force driven Burgers'
equation if cooling and heating are ignored.  Just how well
the dynamics of the
IGM can be described as a Burgers fluid has not been fully
investigated
probably because cooling and heating are essential for a
detailed understanding
of the IGM. Using IGM samples produced by a cosmological
hydrodynamic
simulation in which heating and cooling processes are properly
accounted for,
we show that the IGM velocity field in the nonlinear regime
shows the features of a Burgers fluid, that is, when the Reynolds
number is high, the velocity field consists of an ensemble of
shocks. Consequently, (1) the IGM velocity $v$ is generally
smaller than that of dark matter; (2) for the smoothed field, the
IGM velocity shows tight correlation with dark matter given by $v
\simeq s v_{dm}$,  with $s<1$, such that the lower the redshift,
the smaller $s$; (3) the velocity PDFs are asymmetric between
acceleration and deceleration events; (4) the PDF of velocity
difference $\Delta v=v(x+r)-v(x)$ satisfies the scaling relation
for a Burgers fluid, i.e., $P(\Delta v)=(1/r^y)F(\Delta v/r^y)$.
We find the scaling function and parameters for the IGM which are
applicable
to the entire scale range of the samples (0.26 - 8 h$^{-1}$
Mpc). These properties show that the similarity mapping between
the IGM and dark matter is violated on scales much larger
than the Jeans length of the IGM.

\end{abstract}


\section{Introduction}

The evolution of large scale structure in the universe is largely
governed by the gravitational clustering of dark matter.  The
collapsing of the dark matter halos form the potential wells that
trap baryonic gas, or the intergalactic medium (IGM), and
form light--emitting objects. Clearly understanding the dynamics
of the IGM embedded in the underlying dark matter field is essential in
large scale structure study.

The cosmic baryonic gas is a Navier-Stokes fluid. Unlike an
ordinary fluid, the
dynamic behavior of the IGM is dominated by the growth modes of
dark matter.
It has long been recognized that the growth mode dynamics of
cosmic baryonic
gas can be approximately described by the random-force-driven
Burgers'
equation. Zeldovich's pancake theory and the adhesion model
first indicated
that the non-linear evolution of the gravitational clustering
of cosmic matter
could be sketched by a Burgers' equation (Gurbatov, Saichev, \&
Shandarin 1989; Vergassola et al. 1994). By considering
cosmic matter as dissipative, Berera \& Fang (1994) showed that
the dynamical equations of the velocity fields of cosmic matter are
essentially a variant of the random-force-driven Burgers equation
or the KPZ equation (Kardar et al 1986). Later, with a
two-component (dark matter and IGM) generalization of the adhesion
model, it was found that the velocity potential of the baryonic
matter is described by the Burgers equation driven by the
gravitational potential of dark matter (Jones, 1999; Matarrese and
Mohayaee 2002).

Observations also reveal that the IGM shows features that hint at
a field governed by the dynamics of Burgers' equation. For
example, the transmitted flux of QSO's Ly$\alpha$ absorption is
found to be significantly intermittent and its probability
distribution functions (PDF) are long tailed and probably
lognormal (Jamkhedkar et al. 2000, Pando et al. 2002, Feng et al.
2003, Jamkhedkar et al. 2003). This is consistent with the
intermittent spatial and temporal behavior of a developed Burgers
fluid (Polyakov 1995, Bouchaud, M\'ezard \& Parisi 1995, Gurarie
\& Migdal 1996, Balkovsky et al. 1997, Yakhot 1998, Frisch, Bec \&
Villone 1999). Moreover, if the random-force in the Burgers'
equation for the baryonic matter arises from the gravity of dark
matter, the statistical properties of the baryonic matter in the
non-linear regime will decouple from those of the underlying dark
matter. This mechanism leads to a breaking of similarity between
the IGM and dark matter on scales larger than the Jeans length
(He et al. 2005). This large scale dis--similarity has also been
found from observations
of the entropy floor (Ponman, et al. 1999), X-ray
luminosity-temperature relation of galactic clusters; the baryon
fraction of clusters (Ettori \& Fabian 1999, Sand et al. 2003);
and the multi-phase IGM temperature field (Zhan et al.
2003) etc.

However, there is a difference between the theoretical models and
the observations. In all the above-mentioned theoretical
approaches, heating and cooling processes of the IGM are ignored.
That is, the dynamical equation of the
growth modes of the IGM {\it without} heating and cooling can be
approximated by a Burgers' equation. However, heating and cooling
are essential for understanding all details of the IGM. Therefore, an
important question is how well the dynamics of IGM can be
described as a Burgers
fluid that has undergone significant heating and cooling evolution.
To investigate this question, we focus on the velocity field or
velocity
potential because predictions based on Burgers dynamics
directly concern the velocity and velocity potential. We use
samples of
the IGM velocity field produced by cosmological hydrodynamical
simulations in
which all heating and cooling processes are properly accounted
for and
investigate whether the IGM displays the common features of a
Burgers' fluid.
This problem is important not only for understanding the non-linear
behavior of baryonic gas, but also for the theory of a Burgers'
fluid.

The paper is organized into the following sections: \S 2 reviews
the derivation of the Burgers' equations and its basic predictions
for the baryonic gas. \S 3 gives a brief description of the
hydrodynamic cosmological simulations.  \S 4
gives the statistical analysis of the velocity field of the
baryonic gas. Finally, in \S 5, we will present our conclusions
followed by a discussion of the results. The velocity field in the
discrete
wavelet (DWT) representation is given in the Appendix.

\section{Cosmic baryonic gas velocity field }

\subsection{Burgers' equation for cosmic baryonic gas}

The derivation of Burgers' equation for cosmic baryonic gas
has been addressed in Feng et al (2004), He et al. (2004) and
Pando et al. (2004). In summary, the baryonic gas is assumed to be
an ideal fluid satisfying the hydrodynamic equations
\begin{equation}
\frac{\partial \delta}{\partial t} +
  \frac{1}{a}\nabla \cdot (1+\delta) {\bf v}=0
\end{equation}
\begin{equation}
\frac{\partial a{\bf v}}{\partial t}+
 ({\bf v}\cdot \nabla){\bf v}=
-\frac{1}{\rho}\nabla p - \nabla \phi
\end{equation}
\begin{equation}
\frac{\partial {\cal E}}{\partial t}+5\frac{\dot{a}}{a}{\cal E}+
  \frac{1}{a}\nabla\cdot ({\cal E}{\bf v})=
   -\frac{1}{a}\nabla\cdot(p{\bf v})-
   \frac{1}{a}\rho_{igm}{\bf v}\cdot \nabla\phi- \Lambda_{rad},
\end{equation}
where $\rho$, ${\bf v}$, ${\cal E}$ and $p$ are, respectively,
the mass density, peculiar velocity, energy density and pressure
of the gas. The term $\Lambda_{rad}$ in Eq.(3) is given by
radiative heating and cooling per unit volume of the baryonic gas.
The
gravitational potential $\phi$ in eqs.(2) and (3) is given by
\begin{equation}
\nabla^2 \phi = 4\pi G a^2\bar{\rho}_{dm}\delta_{dm},
\end{equation}
where $\bar{\rho}_{dm}(t)$ and $\delta_{dm}$, are respectively,
the mean
mass density and density contrast of the perturbations of dark
matter.
Here we assume that the gravitational potential $\phi$ is only
given by the dark matter mass perturbation.

To sketch the evolution of gravitational clustering we only consider
the case where all thermal processes are approximated by the
polytropic relations $p \propto \rho^{\gamma}$,
$T \propto \rho^{\gamma-1}$, or $T =T_0(1+\delta)^{\gamma-1}$, where
$\delta({\bf x},t) =[\rho({\bf x},t)-\bar{\rho}(t)]/\bar{\rho}(t)$
is the baryon mass density perturbation. The momentum equation is then
\begin{equation}
\frac{\partial a{\bf v}}{\partial t}+
 ({\bf v}\cdot \nabla){\bf v}=
  -\frac{\gamma k_B T}{\mu m_p} \frac{\nabla \delta}{(1+\delta)}
  - \nabla \phi
\end{equation}
where the parameter $\mu$ is the mean molecular weight of the IGM
particles, and $m_p$ the proton mass. Using the linear approximation
for $\delta$ in the temperature-dependent term of eq.(5), we have
\begin{equation}
\frac{\partial a{\bf v}}{\partial t}+
 ({\bf v}\cdot \nabla){\bf v}=
  - \frac{\nu}{a}\nabla^2 {\bf v}
  - \nabla \phi
\end{equation}
where the coefficient $\nu$ is given by
\begin{equation}
\nu=\frac{\gamma k_BT_0}{\mu m_p (d \ln D(t)/dt)},
\end{equation}
in which $D(t)$ describes the linear growth behavior. The term with
$\nu$ in eq.(6) describes a diffusion characterized by the Jeans
length $k_J^2=(a^2/t^2)(\mu m_p/\gamma k_BT_0)$.

To understand cosmic large scale structure formation,
only the growth modes are of interest. In this case, the velocity
field of the perturbations is irrotational and we can define a
velocity potential by
\begin{equation}
{\bf v}=- \frac {1}{a}\nabla \varphi.
\end{equation}
Substituting eq.(8) into eq.(6), we have
\begin{equation}
\frac{\partial \varphi}{\partial t}-
\frac{1}{2a^2}(\nabla \varphi)^2 -
\frac{\nu}{a^2}\nabla^2 \varphi =\phi.
\end{equation}
Equation (9) is the stochastic-force-driven Burgers' equations or
the KPZ equations (Kardar, Parisi \& Zhang 1986; Barab\'asi \&
Stanley 1995). This is the simplest nonlinear Langevin equation for
growth modes. The second term on the l.h.s. is from convection
which is the lowest-order term for nonlinear growth of clustering.
The third on the l.h.s. describes relaxation of the clustering by
diffusion. The term on the r.h.s is the gravitational potential of
dark matter, which provides the initially random perturbations.
Clustered
structures will develop from the initial seeds via the competition
of the
convection and diffusion. The Burgers' equation has been extensively
applied in the non-linear dynamics of growing structures from a
smoothed surface driven by random forces (e.g. Vicsek 1992). Eq. (9)
shows that structure formation of baryonic gas in the universe can
be thought of as the evolution of surface roughening, i.e., an
initially smoothed distribution evolves into a wrinkled one.

\subsection{Baryonic gas as a Burgers' fluid}

Much theoretical work on Burgers fluids has been done in the
hydrodynamic community in the past decade (Polyakov 1995; Bouchaud
et al. 1995; Yakhot 1998; L\"assig, 2000; Davoudi et al. 2001).
Here we summarize the salient points necessary for testing whether
cosmic
baryonic gas behaves as Burgers' fluid.

In the Burgers' equation, there are two scales: the dissipation length
or the Jeans length $1/k_J$ and the correlation length of the random
force, $r_c$. The intensity of the random force $\phi$ can
be quantified by the density contrast of dark matter $\delta_{dm}$. The
basic feature of the Burgers equation is that turbulence will developed
in the fluid if the following condition holds (e.g.\, L\"assig 2000;
Feng et al 2003)
\begin{equation}
(k_Jr_c)^{2/3}\langle \delta_{dm}^2\rangle^{1/3} > 1.
\end{equation}
This condition corresponds to the system being
in the non-linear regime of the dark matter field.  In linear regime,
$\delta_{dm}\ll 1$, and therefore, perturbations on scale
$r_c > 1/k_J$ will not cause Burgers' turbulence in the IGM.
In the nonlinear regime, $\delta_{dm}> 1$, and Burgers' turbulence
on scales larger the Jeans length can develop. Since the dark matter
field becomes nonlinear first on small scales and then on large scales,
one can expect that Burgers' turbulence in IGM develops
from small to large scales

Burgers turbulence is qualitatively different from turbulence
described by the Navier-Stokes equations. The latter generally
consists of vortices on various scales, while the former is
a collection of shocks. These features arise because for growth modes,
the fluid is potential and the velocity field is irrotational. If
$\nu \rightarrow 0$, the velocity field in the Burger prescription
acquires
singularities due to the discontinuities caused by strong shocks.
The nonlinear feature of the baryonic gas velocity field with
fully developed Burgers turbulence can be understood as a field
consisting of these shocks (L\"assig 2000).

Since dark matter is not affected by Burgers turbulence, the IGM
velocity field dynamically decouples from the dark matter field
on scales larger than the Jeans length when Burgers turbulence develops.
Jeans diffusion will also lead to the decoupling between the
velocity fields of IGM ${\bf v}$ and dark matter ${\bf v}_{dm}$.
However, the decoupling given by the Jeans diffusion is very different
from that given by the Burgers turbulence shocks. For the former,
the distribution ${\bf v}$ and
$\Delta {\bf v}\equiv {\bf v}({\bf x +r})- {\bf v}({\bf x-r})$ is
symmetric with respect to the
transformation ${\bf v}_{dm} \rightarrow -{\bf v}_{dm}$, i.e.,
the velocity PDF with $\Delta {\bf v}\cdot{\bf v}_{dm}/|{\bf v}_{dm}|>0$
(acceleration in the direction of ${\bf v}_{dm}/|{\bf v}_{dm}|$) is the
same as the PDF with $\Delta {\bf v}\cdot{\bf v}_{dm}/|{\bf v}_{dm}|<0$
(deceleration in the direction of ${\bf v}_{dm}/|{\bf v}_{dm}|$). For
Burgers turbulence,
the shocks consist of an acceleration ramp followed by a rapid
deceleration.
Therefore, the IGM is not symmetric between the sections accelerating
and decelerating. Since the
acceleration is due to the gravity of dark matter, the acceleration
is generally in the direction of ${\bf v}_{dm}$. Thus, the IGM velocity
field ${\bf v}$ is asymmetric with respect to the transformation
${\bf v}_{dm} \rightarrow -{\bf v}_{dm}$. The PDF of $\Delta {\bf v}$ with
$\Delta {\bf v}\cdot{\bf v}_{dm}/|{\bf v}_{dm}|>0$ will not be the
same as the PDF of
$\Delta {\bf v}$ with $\Delta{\bf v}\cdot{\bf v}_{dm}/|{\bf v}_{dm}|<0$.

If $\nu =0$, Burgers equation (9) is scale free when the random
force $\phi$ is scale-free. In this case, the velocity field of the
Burgers fluid is self-similar and the PDF of the velocity
difference $\Delta {\bf v}$ is scale-invariant. The velocity
difference can be effectively measured by the longitudinal velocity
difference defined as
$\Delta v_r\equiv v_{\|}({\bf r}_1, t)-v_{\|}({\bf r}_2, t)$, where
$v_{\|}\equiv {\bf v}\cdot ({\bf r}_1-{\bf r}_2)/r$ and
$r= |{\bf r}_1-{\bf r}_2|$. $\Delta v_r$ measures the acceleration
and deceleration along direction ${\bf r}_1-{\bf r}_2)/r$. A detailed
analysis shows that for a self-similar velocity field, the PDF of
$\Delta v_r$ is (e.g. Davoudi et al 2001)
\begin{equation}
P(\Delta v_r,r)=\frac{1}{r^y}F\left (\frac{\Delta v_r}{r^y}\right),
\end{equation}
where parameter $y$ depends on the statistical properties of the random
force.

Summarizing, if the IGM can be described as a Burgers' fluid, its
velocity
field should show the following properties: 1.) the peculiar velocity
of the IGM at a given point will generally be lower than that of
dark matter
at the same point, 2.) the PDF of
$\Delta {\bf v}_r\cdot{\bf v}_{dm}/|{\bf v}_{dm}|>0$ will not be the
same as the PDF of
$\Delta {\bf v}_r\cdot{\bf v}_{dm}/|{\bf v}_{dm}|<0$, and
3.) The PDF of $\Delta {\bf v}_r$ scales. We will test for these
properties in the following sections.

\section{Cosmological hydrodynamic simulations}

Burgers equation (9) is obtained under the assumptions that 1) decaying
modes are ignored; 2) the cooling and heating of the IGM are ignored
and replaced by a polytropic equation of state, and 3) the diffusion
term is given in the linear approximation. Moreover, the driving force
in the Burgers equation for the baryonic gas comes only from the dark
matter gravity, which itself undergoes evolution. It is not at all clear
that the features of a Burgers fluid occur in the IGM for
which cooling and heating processes are essential. To answer this
question,
we determine whether samples produced by a fully cosmological
hydrodynamical simulation have the properties listed at the end of last
section.

\subsection{The WENO algorithm}

In order to capture the Burgers fluid features, simulations need to
be effective in capturing shocks and discontinuities. Hence we do not
use a Lagrangian approach such as the smoothed particle hydrodynamic
(SPH) algorithm. The reason is the SPH scheme does not handle shocks or
discontinuities well because the nature of SPH is to smooth the
fields (e.g. Borve, Omang, \& Trulsen, 2001; Omang, Borve, \&
Trulsen 2003.) Consequently, Burgers fluid features are overlooked.

We instead take a Eulerian approach. A well-known problem with the
Eulerian
algorithm are the unphysical oscillations near a discontinuity. An
effective method to reduce the spurious oscillations is given by
designed
limiters, such as the Total-Variation Diminishing (TVD) method (Harten
1987 et al.) or the piecewise parabolic method (PPM) (Collella \&
Woodward 1984).
However, the accuracy in the TVD method degenerates to first order
near smooth extrema (Godlewski \& Raviart 1996). This problem will
cause errors in calculating the temperature and entropy changes
because they are determined by the difference of the thermal
energy $P/(\gamma-1)$ on two sides of the shock. When the Mach
number of gas is high, the thermal energy $P/(\gamma-1)$ is very
small compared to the kinetic energy $\rho {\bf v}^2/2$. To overcome
this problem, the Essentially Non-Oscillatory (ENO) algorithm
and Weighted Essentially Non-Oscillatory (WENO) were proposed
(Harten et al. 1987; Shu 1998; Fedkiw, Sapiro \& Shu 2003; Shu 2003).
They can simultaneously provide a high order precision for both the
smooth part of the solution and the sharp shock transitions(Liu,
Osher, \& Chan 1994; Jiang \& Shu 1996).

WENO has been successfully applied to hydrodynamic problems containing
shocks and complex structures. WENO has also been used to study
astrophysical hydrodynamics, including stellar atmospheres (del
Zanna, Velli \& Londrillo 1994 ), high Reynolds number compressible
flows with supernova (Zhang et al. 2003), and high Mach number
astrophysical jets (Carrillo et al. 2003). In the context of
cosmological applications, WENO has proved to be especially
adept at handling the Burgers' equation (Shu 1999). Recently, a
hybrid hydrodynamic/N-body code based on the WENO scheme was
developed. It has passed typical reliability tests including the
Sedov blast wave and the formation of the Zeldovich pancake (Feng, Shu
\& Zhang
2004). This code has successfully produced QSO Ly$\alpha$ transmitted
flux
samples, including the high resolution sample HP1700+6416
(Feng, Pando \& Fang 2003). The statistical features of these
samples are in
good agreement with observed features not only on second order
measures, like the power spectrum, but also up to orders as high as
eighth for the
intermittent behavior.

\subsection{Samples}

For the purpose of this paper, we run the hybrid
$N$-body/hydrodynamic code to trace the cosmic evolution of the
coupled system of both dark matter and baryonic gas in a flat low
density CDM model ($\Lambda$CDM), which is specified by the
cosmological parameters
$(\Omega_m,\Omega_{\Lambda},h,\sigma_8,\Omega_b)=
(0.3,0.7,0.7,0.9,0.026)$. The baryon fraction is fixed using the
constraint from primordial nucleosynthesis as
$\Omega_b=0.0125h^{-2}$ (Olive et al. 1991). The value of
$\Omega_b$ is lower than the current result from WMAP. However, since we
we need to refer our previous results (Feng et. al. 2003, Pando et. al. 2004, 
He et. al. 2004), which are based on the simulation samples produced before 
WMAP, we use the older value. The baryon fraction is not a parameter that 
enters into Burgers' equation, and therefore Burgers' fluids features do 
not seriously depend on the baryon fraction once the gravitation
of the baryon gas is negligible in comparison to that of dark matter.

The simulations are performed in a periodic, cubic box of size
$L^3=$25 h$^{-1}$Mpc with a 192$^3$ grid and an equal number of
dark matter particles. The size of the grid is
$25/192=33/2^8=0.129$ h$^{-1}$ Mpc. The thickness of the Burgers'
shock is on the order of the dissipation length, i.e., the Jeans
length, which is in the range $\sim 0.1 - 0.3$ h$^{-1}$ Mpc for
redshifts $z<4$ (Bi et al. 2003). Therefore, the resolution of our
simulation is enough to capture shocks. For strong shocks, the
thickness of the shock is less than the dissipation length.
However, strong shocks are generally associated with high
temperature regions (i.e. $\gg 10^4$ K), and the local Jeans
length is larger. Therefore, the resolution is still less than the
Jeans length. In Feng et al 2003, we used two samples with
different resolution. The major statistical results were found to
be independent of the resolutions.

The simulations start at a redshift
$z=49$ and the results are recorded at redshifts $z=$6.0, 4.0, 3.0,
2.0, 1.0, 0.5 and 0.0. The time step is chosen by the minimum value
among
the following three time scales. The first is from the Courant
condition.  The second time scale is imposed by
cosmic expansion which requires that $\Delta a /a <0.02$ within a
single
time step. The last time scale comes from the requirement that a
particle
moves not more than a fixed fraction of the cell size.

Atomic processes, including ionization, radiative cooling and
heating in a plasma of hydrogen and helium of primordial composition
($X=0.76$, $Y=0.24$) are modeled in the same way as in Bahcall \&
Cen (1992).
Processes such as star formation, feedback due to SN and AGN
activities
have not yet been taken into account.  A uniform
UV-background of ionizing photons is assumed to have a power-law
spectrum of the form $J(\nu) =J_{21}\times10^{-21}
(\nu/\nu_{HI})^{-\alpha}$erg s$^{-1}$cm$^{-2}$sr$^{-1}$Hz$^{-1}$,
with parameter $J_{21}=1.0$ and $\alpha=1$. The photoionizing flux
is suddenly switched on at $z > 10$ to heat the gas and reionize
the universe.

For our work here, we randomly sampled 192 data points along 500
one-dimensional directions ($500\times 192 = 9.6 \times 10^4$
total points.) At each point, the mass density $\rho_{dm}$ and
peculiar velocity component along the line of sight $v_{dm}$ of
dark matter, and mass density $\rho_{igm}$, peculiar velocity
component along the line of sight $v$ and temperature $T$ of the
baryonic gas are recorded. At times during our analysis it will be necessarry
to smooth the  distributions on scales of $33/2^j$ h$^{-1}$ Mpc, where 
$j=3, 4, 5, 6, 7$. 

\section{Statistical analysis of velocity fields}

\subsection{Relation between the IGM and dark matter velocity fields}

In the linear growth regime, the IGM velocity follows the dark
matter velocity point by point, i.e., ${\bf v}({\bf x},t)={\bf
v}_{dm}({\bf x},t)$ (Bi, B\"orner, \& Chu 1993; Fang et al. 1993;
Bi 1993, Nusser 2000.) Burgers' turbulence will lead to a
deviation of ${\bf v}({\bf x},t)$ from ${\bf v}_{dm}({\bf x},t)$.
Figure 1 gives the relation between the velocities of the IGM and
dark matter at each grid point of the samples at redshifts 6, 4,
2, 1, 0.5 and 0, where the data is smoothed on scale $33/2^7$
h$^{-1}$ Mpc. As expected, the data points are scattered around
the diagonal line $v =v_{dm}$. At high redshift $z=6$, the scatter
is smaller, reflecting the fact that the fields are still in the
quasi-linear regime. At $z < 1$, the scatter becomes significant.
The scatter at low redshift cannot be explained by the Jeans
diffusion or random noise as Figure 1 shows a remarkable
butterfly-like configuration, i.e., $v$ mostly lies $0< v <
v_{dm}$ if $v_{dm}>0$, while it lies $0> v
> v_{dm}$ when $v_{dm}<0$. In other words, the peculiar velocity
of IGM $|v|$ is generally less than $|v_{dm}|$. As mentioned in \S
2.2, shocks naturally yield the butterfly-like configuration
because shocks always lead to a deceleration of the gas with
respect to the underlying dark matter.

Figure 1 also shows an envelope of the scatter at about
$v \simeq v_{dm}/4$. This means, the velocity decrease due to shocks
generally is not larger than a factor of 4. This factor is consistent
with the shock theory of a polytropic gas (Landau \& Lifshitz 1959).
The strongest shocks of a polytropic gas lead to an decrease of the
velocity by a factor of $\sim (\gamma-1)/(\gamma+1)$. For these samples
the value of $\gamma$ is found to be 5/3 (He at al. 2004). Therefore,
the velocity decrease should not be larger than a factor
$(\gamma+1)/(\gamma-1)=4$.

Figure 2 shows the  relation beteen $v$ and $v_{dm}$ at $z=0$ for
the density ranges: (1) low density $ \rho_{dm} <1$, (2) moderate
density $1 < \rho_{dm} <5$, (3) high density $5 < \rho_{dm}$. We see
that even in the low and moderate density region, the scatter of $v$ is
still butterfly-like. This means that the velocity fields are
dominated  by shock structures not only around massive halos, but
also in moderate density areas. Burgers turbulence consists of shocks
regardless of whether the density is high.

Figure 3 shows again the relation between $v$ and $v_{dm}$, but
$v$ and $v_{dm}$ are now smoothed on scales $d=
33.3/2^j$ h$^{-1}$ Mpc. We see that the butterfly feature is
weaker with increasingly smooth scales. However, this does not
mean that the deviation of $v$ from $v_{dm}$ is smaller. On the
contrary, the distribution of $v$ still substantially deviates
from the diagonal line for smoothing scales from  $2$ to 8 h$^{-1}$
Mpc. This happens because when the smoothing scale is larger than
the mean separation of strong shocks, the strong shocks are
localized within each area.  Consequently, the mean velocity in all
regions is always dominated by these strong shocks.

This point can be more clearly seen in Figure 4, which plots the
relation between $v$ and $v_{dm}$ at redshifts 6, 4, 2, 1, 0.5,
and 0 for the field smoothed on scales $d=8$ h$^{-1}$ Mpc. Figure
4 shows that at redshift 6, the velocity distribution basically
follows $v=v_{dm}$ with little scatter. As redshift decreases, the
distribution of $v$ is always tightly correlated with $v_{dm}$,
but the correlation between $v$ and $v_{dm}$ is not the diagonal
line $v=v_{dm}$, but $v=s v_{dm}$ with the slope $s <1$. The
parameter $s$ decreases with redshift, it is 0.9 at $z=6$, and 0.4
at z=0. Essentially, $s$ is a measure of the effect of
strong shocks at various redshift eras. Since Burgers' shocks
are related to the formation of cusps in the underlying dark matter
density field, the shock growth is parallel with the development
of non-linear clustering. Therefore, the deviation of s from 1 is
bigger for lower redshifts.

\subsection{Relations of velocity-density and velocity-temperature}

We next analyze the relation between mass density and velocity.
Figure 5 shows $v$ vs. $\rho_{dm}$, $v$ vs.
$\rho_{igm}$, $v_{dm}$ vs. $\rho_{dm}$ and $v_{dm}$ vs.
$\rho_{igm}$ at each grid point of the samples at redshift $z=0$.
At a given $\rho_{igm}$ or $\rho_{dm}$, the scatter of $v_{dm}$ is
much larger than $v$.

More interesting is the qualitative difference between the scatter
of $v$ and $v_{dm}$. The scatter of the  IGM velocity $v$ has a
clear $\rho_{dm}$(or $\rho_{igm}$)-dependent envelope: the smaller
the mass densities, the smaller the scatter. This is because the
IGM flow is potential. In low density areas, the gravitational
potential does not vary significantly. Bernoulli's equation
predicts that the greatest possible value of fluid velocity for a
potential flow is proportional to the speed of sound (Landau \&
Lifshitz 1959). Since the speed of sound generally increases with
gas density, the maximum IGM velocity, i.e., the velocity at the
$v-\rho_{igm}$ envelope, increases with $\rho_{igm}$. Although
$\rho_{igm}$ and $\rho_{dm}$ are not the same point-by-point, a
large $\rho_{dm}$ on average corresponds to large $\rho$ (Pando et
al 2004). Therefore, the relation $v$ vs. $\rho_{dm}$ shows a
similar envelope. On the other hand, the dark matter velocity
$v_{dm}$ is not affected by the IGM shocks and is not constrained
by the speed of sound of the IGM. There is no density-dependent
envelope for $v_{dm}$.

Figure 6 plots the relation between the IGM velocity $v$ and
temperature $T$ at
each grid point of the samples at redshift $z=0$. The scatter of
the IGM velocity $v$ with respect to $T$ also has a clear
envelope: the smaller the temperature, the smaller the scatter.
This envelope develops for the same reason as above: the IGM
velocity cannot be much larger than the speed sound. However,
figure 6 shows that the envelope increases with $T$ quickly at
$T\leq 3,000$ K, but slowly at $T\geq 3,000$ K. This is probably
because most hydrogen atoms are neutral at $T\leq 3,000$ K and
ionized at $T\geq 3,000$ K. On the other hand, the velocity of dark
matter $v_{dm}$ is not subject to this constrains. Therefore, there is
no envelope in the $v_{dm}$-$T$ scatter distribution.

It should be pointed out that the envelopes shown in
Figures 5 and 6 significantly deviate from the usual relation
$v\propto \sqrt(T)$ for the speed of sound. The reason is that the baryon gas
exists in different regions of $T -- \rho$ space.  That is, the baryon gas is
in multiple phases.  The multiple phases of cosmic
baryon gas has been addressed in He, Feng \& Fang (2004) where it is  shown 
that the $T-\rho$, and $S({\rm entropy})-T$ and $S-\rho$ relations are 
significantly multiply phased either at low or high temperature or density. 
The deviation from the usual relation of the velocity and temperature shown in
Figs 7 and 8 is evidence that the baryon gas is substantially multiple phased.

\subsection{Asymmetry of the PDF of velocity difference}

As mentioned in \S 2.2, the features of a Burgers fluid are
effectively described by the statistics of the longitudinal
velocity difference $\Delta v_r({\bf x})= v_{\|}[{\bf x}+({\bf
r}/2)]-v_{\|}[{\bf x}-({\bf r}/2)]$. This
quantity describes the velocity difference at physical position
${\bf x}$ and on distance $r= |{\bf r}|$. That is,
$\Delta v_r({\bf x})>0$
corresponds to an acceleration event in direction ${\bf
r}$, while $\Delta v_r({\bf x})< 0$ is a deceleration event. More
useful is to define the acceleration and deceleration with respect
to the velocity direction of the underlying dark matter velocity
field. For one-dimensional fields, the events of $\Delta v_r( x)
v_{dm}(x)/|v_{dm}(x)|>0$ correspond to an acceleration with
respect to the dark matter velocity field, while $\Delta v_r( x)
v_{dm}(x)/|v_{dm}(x)|<0$ the deceleration events.

The velocity difference $v(x+r)-v(x-r)$ contains position $x$ and
scale $r$. It is convenient to measure $\Delta v_r(x)$ by the
discrete wavelet transform (DWT) $\Delta v_{j,l}$ (see Appendix).
Moreover, the DWT modes are orthogonal and effectively avoid
false correlations and scalings. The smoothing can also be done
with the DWT, as the scaling functions on various scales are window
functions on that scale. For our 1-dimensional samples, the DWT
variable $\Delta v_{j,l}$ (see Appendix) describes the velocity
difference at the physical position of cell $l$ and on distance
$r=33.3/2^j$ h$^{-1}$ Mpc. With the DWT variables, the
acceleration and deceleration events with respect to the dark
matter field are given by $\Delta v_{j,l} (v_{j,l}/|v_{j,l}|)_{dm}$.

We showed in \S 4.1 that the IGM velocity is generally lower than
the dark matter velocity because of shocks. Shocks cause a
decrease in the velocity in the direction of the gas flow $ v$.
This means the PDF of $\Delta v_{j,l}$ will be asymmetric between
acceleration and deceleration events. Figure 7 plots the PDFs of
$\Delta v_{j,l}$ for samples at $z=0$ and at smoothing scales $j=$
3, 4, 5 and 6. It is clear that the PDFs of acceleration and
deceleration branches of the distribution are asymmetric. For
$j\leq 5$, the PDFs of deceleration events generally have a longer
tail than that of acceleration events, i.e., there are more big
deceleration events than acceleration events. Using a $K-S$ test, we find
that the probability the the acceleration and deceleration branches are
drawn from the same distribution is $0.048$ for $j=4$, and $0.28$
for $j=3$. Therefore, the asymmetry is weaker on larger scale.

Generally, the asymmetric long tail is significant at $|\Delta
v_{j,l}|>20$ km/s. For $|\Delta
v_{j,l}(v_{j,l}/|v_{j,l}|)_{dm}|\leq 20$ km/s, the asymmetry is
weaker, because the velocity difference caused by the Jeans
diffusion is equal to about 20-30 km/s. This is also the reason
that the asymmetry almost disappeared on scales equal to and less
that 0.5 h$^{-1}$ Mpc, or $j=6$. The K-S probability is $0.15$ for
$j=5$ and 0.79 for $j=6$.  The thickness of shocks due to the
Jeans diffusion is on order $\nu/\langle v^2\rangle^{1/2}$, where
$\langle v^2\rangle^{1/2}$ is the {\it rms} of velocity field.
Therefore, the velocity difference $\Delta v_r(x)$ with $r$ less
than $\nu/\langle v^2\rangle^{1/2}$ will not be sensitive to the
shock discontinuity. Moreover, the 1-dimensional samples will also
weaken the asymmetry. Since $v_{dm}$ is a projection of the
3-dimensional ${\bf v}_{dm}$, the asymmetry will be missed for
events perpendicular to the direction of 1-dimension sampling.

Figure 8 shows the PDFs of $\Delta v_{j,l}$ on scale $j=4$, but
for redshifts $z=0, 1.0, 2.0$ and 4.0. This figure shows that the
PDFs is already asymmetric at $z=4$. The number of large
deceleration events is higher for lower redshifts. This yields
the slope $s<1$ in Figure 4 that is smaller for lower redshifts.

\subsection{Scaling of the $\Delta v_{j,l}$ PDFs }

An important property of the Burgers fluid is the self-similarity
of the velocity field as given by eq.(11). A simplified version
of eq.(11) can be obtained as follows. For the IGM field, the
velocity gradient is $(\Delta v_r)/r$. If the field of the
Burgers' turbulence shocks is self-similar, the PDF of $(\Delta
v_r)/r$, $F[(\Delta v_r)/r]$, should be scale-invariant. In this
case, the PDF of $\Delta v_r$, $P(\Delta v_r)$, is given by
$P(\Delta v_r)d(\Delta v_r) = F[(\Delta v_r)/r]d(\Delta v_r/r)$.
Thus, we have $P(\Delta v_r)=(1/r)F[(\Delta v_r/r)$. This is the
PDF of eq.(11) with $y=1$.

Using DWT variables and noting that the physical scale $r$ is
given by index $j$ as $r=33.33/2^j$, eq.(11) can be rewritten as
\begin{equation}
P(\Delta v_{j,l}) = 2^{yj}F(2^{yj}\Delta v_{j,l}).
\end{equation}
Eq.(12) requires that the dynamics of the velocity difference
$\Delta v_{j,l}$ be scale-free. It is not
immediately clear whether the IGM velocity field is really
scale-invariant because the IGM dynamical equations (1) - (4)
contain many physical scales, such as the Jeans scale; the scales
related to cooling and heating, etc., and these are not necessarily
scale free.

We tried to fit the distribution $P(\Delta v_{j,l})$ with a
scale-invariant function like (12). The first result is shown in
Figure 9, which gives the $P(\Delta v_{j,l})$ against $\Delta
v_{j,l}$ for the deceleration branch for samples at $z=0$. We found
the best scaling function is
\begin{equation}
P(\Delta v_{j,l})
= 0.012\times 2^{0.7 j}
\exp \left
[-\left (2^{0.7 j}\frac{\Delta v_{j,l}}{50 {\rm km/s}}\right )^{0.75}
  \right ]
\end{equation}
The PDF eq.(13) is consistent with eq.(12). The quantity 0.012
is given by the overall normalization and can be absorbed by a
unit transformation. The scaling behavior of the PDF depends on
the two scaling factors $2^{0.7 j}$ and $(2^{0.7 j})^{0.75}$,
and therefore, the scaling is given by two parameters: $y= 0.70$,
and the index 0.75. The factor 50 $km/s$ makes
the exponential dimensionless, but is not involved in the scaling
and has no $j$-dependence.
The scaling function (13) excellently fits PDF
$P(\Delta v_{j,l})$ on scales $j=2$ - 7,
i.e., from the highly nonlinear scale 0.26 h$^{-1}$ Mpc to the
weak nonlinear scale 8 h$^{-1}$ Mpc.

Figure 10 is similar to Figure 9, but for the acceleration branch.
The scaling function is now
\begin{equation}
P(\Delta v_{j,l})
= 0.028\times 2^{0.47 j}
\exp \left
[-\left ( 2^{0.47 j}\frac{\Delta v_{j,l}}{28 {\rm km/s}}\right )^{0.75}
  \right ].
\end{equation}
Similar to eq.(13), the scaling behavior of eq.(14) depends
on two parameters: $y= 0.47$, and the index 0.75. The factor 28
$km/s$ makes the exponential  dimensionless.
Eq.(14) also gives a very good fit to the PDFs from $j=2$ to $j=6$.
The PDFs at $j=7$ shows small deviation from eq.(14). This is probably
because the scale $j=7$, or 0.26
h$^{-1}$ Mpc is comparable with the Jeans length. The parameter
$y=0.47$ in eq.(14) is significantly different than value given in
Eq.(13).  This difference indicates that although both the the
acceleration and deceleration branches scale, the scaling
behavior is not symmetric between the acceleration and
deceleration branches.

Both functions (13) and (14) are consistent with eq.(12).
Burgers' equation predicts that the acceleration and
deceleration branches will have different PDFs. All these results show
that the scale invariant behavior of IGM velocity field is the
same as a Burgers' fluid. According to the general theory of
Burgers' fluids (Davoudi et al 2001), the parameter $y$ should be
dependent on the statistical property of the random force $\phi$
and the relation cannot be found analytically. The simulation box of
size $L^3=$25 h$^{-1}$Mpc is not much larger than 8 h$^{-1}$Mpc, but
the scaling shown in Figures 9 and 10 still gives a good fit on
scale of 8 h$^{-1}$Mpc. The reason for this is the use of the
velocity difference which suppresses the contribution of long wave
length perterbations.

\section{Discussions and conclusions}

The initial mass density fluctuations of the universe are probably
Gaussian with a scale-free power spectrum. Therefore, in the
linear regime the dark matter and IGM mass and velocity fields are
self-similar. In the non-linear regime, the dynamical behavior of
the dark matter and IGM are very different even though they are
coupled via gravity. Since the dynamics of the dark matter is
governed only by the gravitational interactions which do not have
a preferred scale, the dark matter should undergo  self-similar
evolution even in the non-linear regime. The
evolution of the IGM seems to be much more complicated because it
sensitively depends on thermal, radiative and other
non-gravitational processes. Consequently, the properties of the
Burgers' equation of the dark matter field cannot simply be
transferred to the IGM (He et al. 2004.)

Nevertheless, we showed that the main dynamical feature of the IGM
velocity field can be explained as a Burgers fluid. The IGM
samples, produced by a full hydrodynamic simulation in which
thermal processes are involved, show all features of a Burgers
fluid. This is probably because the scales considered are
larger than the Jeans length, but smaller than the scale at
which the nonlinear evolution is onset. In this range the IGM is
nonlinear, but not sensitive to the scales of thermal processes.
Therefore, it can also be approximately described as a Burgers'
fluid.

However, as a Burgers' fluid, the IGM velocity field is not a
simple mirror of the underlying dark matter velocity field. For
instance the PDFs of the velocity difference of dark matter is
lognormal (Yang et al. 2001) and significantly different from
eqs.(13) and (14). The PDF of the velocity difference of dark
matter is symmetric between the acceleration and deceleration
branches with respect to the velocity field of dark matter, but
asymmetric for the IGM. All these differences come from shocks
which do not develop in the dark matter. Non-linear evolution will
inevitably lead to a dynamical break of the similarity between the
dark matter and gas.

\acknowledgments

Ping He acknowledges support from the World Laboratory scholarship.
LLF acknowledges support from the National Science Foundation of China
(NSFC) and National Key Basic Research Science Foundation. This work is
partially supported by the National Natural Science Foundation of China
(10025313) and the National Key Basic Research Science Foundation of
China (NKBRSF G19990752).

\appendix

\section{The DWT variables of velocity field}

For the details of the mathematical properties of the DWT refer to
Mallat (1989a,b), Meyer (1992), Daubechies (1992), and for
physical applications, refer to Fang \& Thews (1998).  For us
here, the most important properties are 1.) orthogonality, 2.)
completeness, and 3.) locality in both scale ($r$) and physical
position ($x$). Wavelets with compactly supported basis are an
excellent means to analyze the velocity fields. Among the
compactly supported orthogonal wavelets, the Daubechies family of
wavelets are easy to implement.

To simplify the notation, we consider a 1-D velocity field $v(x)$
on spatial range $L$. It is straightforward to generalize to 3-D
fields. In DWT analysis, the space $L$ is chopped into $2^j$
segments labeled by $l=0,1,...2^j-1$. Each of the segments has
size $L/2^j$. The index $j$ is a positive integer which represents
scale $L/2^j$. The index $l$ gives position and corresponds to
spatial range $lL/2^j < x < (l+1)L/2^j$.

DWT analysis  uses two functions, the scaling functions
$\phi_{j,l}(x)=(2^j/L)^{1/2}\phi(2^j/L-l)$, and wavelets
$\psi_{j,l}(x)=(2^j/L)^{1/2}\psi(2^j/L-l)$. The scaling
functions play the role of window function.  They are used to
calculate the mean field in the segment $l$. The wavelets
$\psi_{j,l}(x)$ capture the difference between
the mean fields at space ranges $lL/2^j < x < (l+1/2)L/2^j$ and
$(l+1/2)L/2^j < x < (l+1)L/2^j$.

The scaling functions and wavelets $\psi_{j,l}(x)$ satisfy the
orthogonal relations
\begin{equation}
\int \phi_{j,l}(x)\phi_{j,l'}(x)dx= \delta_{l,l'},
\end{equation}
\begin{equation}
\int \psi_{j,l}(x)\psi_{j',l'}(x)dx=\delta_{j,j'} \delta_{l,l'},
\end{equation}
\begin{equation}
\int\phi_{j,l}(x)\psi_{j',l'}(x)dx =0, \ \ \ \mbox{if $j'\geq j$}.
\end{equation}

With these properties, a 1-D random field $v(x)$ can be decomposed into
\begin{equation}
v(x) = v^{j}(x) + \sum_{j'=j}^{\infty} \sum_{l=0}^{2^{j'}-1}
  \tilde{\epsilon}_{j',l} \psi_{j',l}(x),
\end{equation}
where
\begin{equation}
v^j(x)=\sum_{l=0}^{2^j-1}\epsilon_{j,l}\phi_{j,l}(x).
\end{equation}
The scaling function coefficient (SFC) $\epsilon_{j,l}$ and the
wavelet function coefficient (WFC), $\tilde{\epsilon}_{j,l}$
are given by
\begin{equation}
\epsilon_{j,l} =\int v(x)\phi_{j,l}(x)dx,
\end{equation}
and
\begin{equation}
\tilde{\epsilon}_{j,l}=\int v(x)\psi_{j,l}(x)dx,
\end{equation}
respectively. The SFC $\epsilon_{j,l}$ measure the mean
of $v(x)$ in the segment $l$, while the WFC
$\tilde{\epsilon}_{j,l}$ measures the fluctuations (or difference)
of field $v(x)$ at $l$ on scale $j$.

The first term on the r.h.s. of eq.(A4), $v^{j}(x)$, is the field
$v(x)$ smoothed on the scale $j$, while the second term contains
all information on scales $\geq j$. Because of the orthogonality,
the decomposition between the scales of $ <j$ (first
term) and $\geq j$ (second term) in eq.(A4) is unambiguous.

Since scale $r$, and position $x$, correspond to, respectively,
$j$, $l$,
and $\phi_{j,l}(x')$ is window filter on scale $r$ at position $x$,
the DWT variables of velocity field are given by
\begin{equation}
  v_{j,l} = \frac{\int v(x)\phi_{j,l}(x) dx}
                 {\int \phi_{j,l}(x) dx}.
\end{equation}
$v_{j,l}$ is the mean
velocity in the spatial range $lL/2^j < x < (l+1)L/2^j$.
Similarly, the DWT counterpart of the velocity difference is
\begin{equation}
\Delta v_{j,l} = \frac{\int v(x)\psi_{j,l}(x) dx}
                 {\int \phi_{j,l}(x) dx}.
\end{equation}
$\Delta v_{j,l}$ is the difference between the mean
velocities of spatial ranges $lL/2^j < x < (l+1/2)L/2^j$ and
$(l+1/2)L/2^j < x < (l+1)L/2^j$.

$v_{j,l}$ and $\Delta v_{j,l}$ are the variables of the velocity
field $v(x)$ in the DWT representation. These variables give a
complete description of the field $v(x)$ without a loss of
information. The orthogonality of scaling functions and wavelets
insure that the decomposition does not cause false correlation
among these variables.

\clearpage

\figcaption[figure1.eps]{$v$ vs. $v_{dm}$ at each grid point
of samples at redshifts 6, 4, 2, 1, 0.5 and 0. The two solid lines
are, respectively, $v=v_{dm}$, and $v=v_{dm}/4$.
\label{fig1}}

\figcaption[figure2.eps]{$v$ and $v_{dm}$ at $z=0$ for different
density ranges: (1) (1) $\rho_{dm}<1$ (top), (2)
$1 < \rho_{dm} <5$ (middle), and (3) $\rho_{dm} > 5$ (bottom).
 \label{fig2}}

\figcaption[figure3.eps]{$v$ and $v_{dm}$ for sample $z=0$. The
velocity field is smoothed on scales  $d= 33.3/2^j$ h$^{-1}$ Mpc,
and $j=2, 4, 6, 8$.
 \label{fig3}}

\figcaption[figure4.eps]{$v$ and $v_{dm}$ for samples at $z=$ 6,
4, 2, 1, 0.5, 0, and smoothed on scale $8$ h$^{-1}$ Mpc. The solid
line is $v=sv_{dm}$. The $v$-$v_{dm}$ relation is fitted with line
$v=sv_{dm}$ (dashed line). The parameter $s$ is shown in each
panel.
 \label{fig4}}

\figcaption[figure5.eps]{The relation of $v$ vs. $\rho_{dm}$ (top
left); $v$ vs. $\rho_{igm}$ (top right); $v_{dm}$ vs. $\rho_{dm}$
(bottom left), and $v_{dm}$ vs. $\rho_{igm}$ (bottom right) at
each grid point of the samples at redshift $z=0$. The envelop of
the $v$-$\rho_{dm}$ distribution is given by
$v=[\log(\rho_{dm}/0.001)]^b$, where $b=2.9$ (dotted); 3.0
(dashed) and 3.1 (solid).
 \label{fig5}}

\figcaption[figure6.eps]{The relation of $v$ vs. temperature $T$
(top) and $v_{dm}$ vs. temperature $T$ (bottom) at each grid point
of the samples at redshift $z=0$. The envelop of $v$-$T$
distribution is given by $v=[g \log(T/100)]^b$, where $b=2.5$;
$g=2.6$ (solid), 3.0 (dashed) and 3.5 (solid).  \label{fig6}}

\figcaption[figure7.eps]{The PDF of velocity difference $\Delta
v_{j,l}$ for acceleration events (solid) and deceleration events
(dotted). The samples are at $z=0$ with smoothing scale
$d=33.3/2^j$ h$^{-1}$ and $j=$ 3, 4, 5 and 6.
 \label{fig7}}

\figcaption[figure8.eps]{The PDF of velocity difference $\Delta
v_{j,l}$ for acceleration events (solid) and deceleration events
(dotted). The samples are at redshifts $z=4$, 2, 1 and 0, with
smooth scale is $j=4$.
 \label{fig8}}

\figcaption[figure9.eps]{The PDF of velocity difference $|\Delta
v_{j,l}|$ of deceleration cases, i.e. $\Delta
v_{j,l}(v_{j,l}/|v_{j,l}|)_{dm}<0$ The samples is for $z=0$. The
scales $r$ in the velocity difference $\Delta v_r=v(x+r)-v(x)$ are
taken to be $j=2, 3, 4, 5$ 6 and 7. \label{fig9}}

\figcaption[figure10.eps]{The PDF of velocity difference $|\Delta
v_{j,l}|$ of acceleration cases, i.e. $\Delta v_{j,l}
(v_{j,l}/|v_{j,l}|)_{dm}>0$ The samples is for $z=0$. The scales
$r$ in the velocity difference $\Delta v_r=v(x+r)-v(x)$ are taken
to be $j=2, 3, 4, 5$ 6 and 7. \label{fig10}}

\end{document}